\documentclass[10pt]{article}
\usepackage{amsmath}
\usepackage{amsfonts}
\usepackage{amssymb}
\usepackage{setspace}
\usepackage{multicol}

\usepackage[left=20mm,right=20mm,top=33.95mm,bottom=33.95mm]{geometry}

\ifx\pdfoutput\@undefined\usepackage[usenames,dvips]{xcolor}
\else\usepackage[usenames,dvipsnames]{xcolor}
\IfFileExists{pdfcolmk.sty}{\usepackage{pdfcolmk}}{} 
\fi
\usepackage[plainpages=false,pdfpagelabels,pagebackref=false,naturalnames=true,hyperindex=true,pdftitle={An Algorithmic Approach to Information and Meaning},pdfauthor={Hector Zenil}]{hyperref}
\hypersetup{colorlinks=true,
urlcolor=Cerulean,linkcolor=BrickRed,citecolor=RoyalBlue,a4paper,
  pdfpagemode=None,
  pdfstartview=FitH}
\usepackage[all]{hypcap}

\newenvironment{myitemize}{
\begin{itemize}
  \setlength{\itemsep}{1pt}
  \setlength{\parskip}{0pt}
  \setlength{\parsep}{0pt}}{\end{itemize}
}

\begin{document}

\begin{multicols}{2}

\title{\Large\textbf{An Algorithmic Approach to\\Information and Meaning:}\\A Formal Framework for a \\Philosophical Discussion\footnote{Presented at the \emph{Interdisciplinary Workshop: Ontological, Epistemological and Methodological Aspects of Computer Science}, Philosophy of Simulation (SimTech Cluster of Excellence), Institute of Philosophy, Faculty of Informatics, University of Stuttgart, Germany, July 7, 2011.}}
\author{\textbf{\large Hector Zenil}\\ \small Institut d'Histoire et de Philosophie des Sciences\\ \small et des Techniques (Paris 1/ENS Ulm/CNRS)\\}
\date{}
\maketitle

\begin{abstract}

I will survey some matters of relevance to a philosophical discussion of information, taking into account developments in algorithmic information theory (AIT). I will propose that meaning is deep in the sense of Bennett's logical depth, and that algorithmic probability may provide the stability needed for a robust algorithmic definition of meaning, one that takes into consideration the interpretation and the recipient's own knowledge encoded in the story attached to a message.\\

\noindent \textbf{Keywords:} information content; meaning; algorithmic probability; algorithmic complexity; logical depth; philosophy of information; information theory.

\end{abstract}

\section{Introduction}

Information can be a cornerstone for interpreting all manner of phenomena, as it can constitute the basis for a description of objects. While it is legitimate to study ideas and concepts related to information in their broadest sense, that the use of information outside formal contexts amounts to misuse cannot and should not be overlooked. It is not unusual to come across surveys and volumes devoted to information (in the larger sense) in which the mathematical discussion does not venture beyond the state of the field as Shannon \cite{shannon} left it some 60 years ago. Recent breakthroughs in the development of information theory in its algorithmic form---both theoretical and empirical developments possessing applications in diverse domains (e.g.  \cite{li,li2,li3,zenilimag})---are often overlooked in the semantic study of information, and it is philosophy and logic (e.g. epistemic temporal logic) that are resorted to in attempting to account for what is said to be the semantic formalism of information. As examples one may cite the work of  \cite{floridi1,floridi2,stanbio}. In the best of cases, algorithmic information theory (AIT) is not given due weight. Its basic definitions are sometimes inaccurately rendered (e.g. the incomplete definition of Bennett's logical depth \cite{bennett} in \cite{nature} (p. 25)).

In \cite{floridi2}, for example, the only reference to AIT as a formal context for the discussion of information content and meaning is a negative one---appearing in van Benthem's contribution (p. 171 \cite{floridi2}). It reads:

\begin{quote}
To me, the idea that one can measure information flow one-dimensionally in terms of a number of bits, or some other measure, seems patently absurd...
\end{quote}

I think this position is misguided. When Descartes transformed the notion of space into an infinite set of ordered numbers (coordinates), he did not deprive the discussion and study of space of any of its interest. On the contrary, he advanced and expanded the philosophical discussion to encompass concepts such as dimension and curvature-- which would not have been possible without the Cartesian intervention. Perhaps this answers the question that Benthem poses immediately after the above remark (p. 171 \cite{floridi2}):

\begin{quote}
But in reality, this quantitative approach is spectacularly more successful, often much more so than anything produced in my world of logic and semantics. Why?
\end{quote}

Accepting a formal framework such as algorithmic complexity for information content does not mean that the philosophical discussion of information will be reduced to a discussion of the numbers involved, as it did not in the case of the philosophy of geometry after Descartes.

 The foundational thesis upon which information theory rests today (derived from Shannon's work) is that information can be reduced to a sequence of symbols. Despite the possibility of legitimate discussions of information on the basis of different foundational hypotheses, in its purely syntactic variant, information theory can be considered in large part achieved by Shannon's theory of communication (see Box 1).

Epistemological discussions are, however, impossible to conceive of in the absence of a notion of semantics. Much work has been done in logic to capture the concept of meaning in a broader and formal sense. However, little or nothing has been done to explain meaning using pure computational models---whether to extend Shannon's work on information, to explain meaning in light of Turing's merging of information and computation, or to explain meaning in light of current developments, as epitomized by the theory of AIT.
 
Semantics is concerned with content. Both the syntactic and semantic components of information theory are concerned with order, the former particularly with the number of symbols and their combinations, while the latter is intimately related to structure. The theory of computation is the context provided by the theory of algorithmic information for discussion of the concept of information. Within this context the description of a message is interpreted in terms of a program. The following sections are an overview of the different formal directions in which information has developed in the last decades. They leave plenty of room for fruitful philosophical discussion, discussion focusing on information per se as well as on its connections to aspects of physical reality.

\section{Communication, diversity and complexity}
 
Shannon's conception of information inherits the pitfalls of probability (see Box 1). Which is to say that one cannot talk about the information content of individual strings. However, misinterpretations have dogged Shannon's information measure from the inception, especially around the use of the term \emph{entropy}, as Shannon himself acknowledged. The problem has been that Shannon's entropy is taken to be a measure of order (or disorder), as if it were a complexity measure (analogous to physical entropy in classical thermodynamics). Shannon acknowledges that his theory is a theory of communication and transmission and not one of information \cite{shannon}. 

Unlike algorithmic complexity (see Box 2) Shannon's entropy does not take into consideration the internal structure of the message. Consider the strings 0101010101 and 0100101100. They both have exactly the same Shannon entropy because the number of occurrences of 1 and of 0 is the same in both strings. As a second example, consider the sentence ``A quick brown fox jumps over the lazy dog" and the scrambled version ``rl y feawhojkouq A vdpegsxioz r cmunotb". They both have the same Shannon entropy value even though clearly one tells us something while the other is nonsensical. Shannon's Entropy basically says that because the members of each of these pairs of messages use about the same number of symbols, one needs a communication channel of about the same size for both members of the pair.\\[-10pt]

\begin{center}
\colorbox{yellow!10}{\fbox{
 \begin{minipage}[t][.315\textheight]{.465\textwidth}
\textbf{Box 1. Shannon's entropy} is defined as a measure of the average information content associated with a random outcome. Formally, Let $d=(p_1,p_2, \ldots ,p_n)$ be a finite discrete probability distribution. That is, suppose $p_k \geq 0$ for $k=1,2, \ldots ,n$ and $\Sigma_{k=1}^n p_k= 1$. The uncertainty concerning a possible outcome with probabilities $p_1, p_2, \ldots, p_n$ is called the entropy of the distribution $P$ and is measured by $H(P)=H(p_1,p_2,\ldots, p_n)$ as introduced by Shannon \cite{shannon} and defined by $H(p_1, p_2, \ldots, p_n) = \Sigma_{k=1}^n p_k \log_2 1/p_k$. It indicates how many bits are required to encode a message in order to send it through a communication channel with minimum capacity. It counts how many different symbols a message has, weighted by their probability distribution. It is therefore a measure of diversity, not of order or disorder.
 \end{minipage}\text{ }}}
 \end{center} \text{\\[10pt]}

That Shannon's measure is computable and easily calculable in practice may account for its frequent and unreasonable application as a complexity measure. The fact that algorithmic complexity is not computable, however, doesn't mean that one cannot approximate it.

 Shannon's approach doesn't help to define information content or meaning. For example, think of a number like $\pi$ which is believed to be normal (that is, that its digits are equally distributed), and therefore has little or no redundancy. The number $\pi$ has no repeating pattern (because it is an irrational number). Lacking a pattern, there is no way to optimize a channel through which to transmit it. $\pi$, however, can be greatly compressed using any of the known briefly describable formulas generating its digits, so that one can send the formula rather than the digits. But this kind of optimization is not within the scope of Shannon's communication theory. Unlike Shannon's treatment of $\pi$, one can think of $\pi$ as a meaningful number because of what it represents: the relationship between any circumference and its diameter. I will argue that meaning can be treated formally, using concepts of AIT to approach these matters.

 The main point of relevance to us made by Shannon when formulating his measure in the context of communication is that in practice a message with no redundancy is more likely to carry information if one assumes one is transmitting more than just random bits. If something is random-looking, then it will usually be considered meaningless. To say that something is meaningful usually implies that one can somehow arrive at a conclusion based on it. Information has meaning only if it has a context, a story behind it. Meaning, in a causal world, is the story attached to a message.

\section{Data $+$ program is message $+$ interpretation}

Among the several contributions made by Alan Turing on the basis of his concept of computational universality is the unification of the concepts of data and program. Turing machines are extremely basic abstract symbol-manipulating devices, which despite their simplicity, can be adapted to simulate the logic of any computer that could possibly be constructed. While one can think of a Turing machine input as data, and a Turing machine rule table as its program, each of them being separate entities, they are in fact interchangeable as a consequence of universal computation, as shown by Turing himself, since for any input $s$ for a Turing machine $M$, one can construct $M^\prime$ with empty input such that $M$ and $M^\prime$ accept the same language, with $M^\prime$ a (universal) Turing machine accepting an encoding of $M$ as input and emulating it for an input $s$ for $M$ in $M^\prime$. 

In other words, one can always embed data as part of the rule table of another machine. The identification of something as data or a program is therefore merely a customary convention and not a fundamental distinction. This is noteworthy because though it may seem that a message is to data as an interpretation is to a program, part of the argument is that the message cannot be formally captured in both computational and semantic terms if the interpretation is not part of it. So even if on the surface we may make a distinction, just as we make a distinction between a message and its recipient in the real world, the difference is not essential. Hence subsuming everything under the heading of 'message' will allow us to define the concept of \emph{meaning} on the basis of the theory of computation and algorithmic information. For example, messages (inputs) for which a Turing machine halts can be taken to be messages that have been understood. It is certain that in the case of a universal Turing machine there will be messages for which the machine will halt and others for which it will not halt, a desirable property insofar as we are concerned with defining meaning. Which is to say that no matter what the meaning of a message, there should always be recipients that are capable of interpreting it and others that are not. That some \emph{understand} a message by halting (i.e., conventionally) doesn't mean, however, that all of them \emph{understand} it in the same way, since each may react to it in a different way--which is desirable insofar as we are seeking to define a concept of subjective meaning. These simple first conventions will allow us to apply several concepts from algorithmic information theory, notably the concepts of conditional complexity, algorithmic probability and logical depth, in order to impart sense and robustness to our algorithmic approach to the concept of \emph{meaning} from a computational perspective.

\section{Information content and meaning}

Nevertheless, Shannon's notion of information makes it clear that information content is subjective (Shannon himself):

\begin{quote}
Frequently the messages have meaning: that is they are referred to or correlated according to some system with certain physical or conceptual entities. These semantic aspects of communication are irrelevant to the engineering problem. The significant aspect is that the actual message is one selected from a set of possible messages. \cite{shannon}.
\end{quote}

\emph{Subjective} doesn't mean, however, that one cannot define information content formally, only that one should include a plausible interpretation in the definition, a point we will explore in the next section.

 Shannon's contribution is seminal in that he defined the bit as the basic unit of information, as do our best current theories of informational complexity. Any sequence of symbols can be translated into a binary sequence, thereby preserving the original content, as it can be translated back and forth from the original to the binary and vice versa. Shannon's information theory approaches information syntactically: whether and how much information (rather than what information) is conveyed. And as a physical phenomenon: the basic idea is to make the communication channel more efficient.

 Unlike Shannon's entropy, algorithmic complexity considers individual objects independent of any probability distribution. For example, different initial segments of the Fibonacci sequence have each a different entropy. Each of the 10 segments having 10 more values of the Fibonacci sequence correspond to the following sequence of Shannon's entropy values: 2.163, 2.926, 3.354, 3.654, 3.88, 4.071, 4.228, 4.36, 4.484, 4.591-- simply because the longer the sequence of numbers, the more bits are required to encode them as such through a communication channel. Yet one can decide that in order to recover the same message at about the same size, it would be preferable to send the formula together with the number of Fibonacci numbers to generate. The sequence can be expressed as $F(n) = F(n-1) + F(n-2)$ for $n=\{3,4,5,...\}$, $F(1)=1$, $F(2)=1$. Which as a formula has a low and constant algorithmic complexity for any segment of the Fibonacci sequence. This example reveals that this is a significant difference: two messages of arbitrarily different algorithmic complexity can have the same Shannon entropy. Shannon's entropy represents an absolute limit on the best possible lossless compression of the algorithmic complexity, i.e. $C(s)$ (program-size algorithmic complexity) cannot be smaller than $H(s)$ (Shannon's entropy). However, the Fibonacci sequence as a message has the same meaning for algorithmic complexity, again another desirable feature of algorithmic complexity (and clearly one that would be unavailable were the problem to be approached through Shannon's entropy). Further information about Shannon's entropy in comparison to algorithmic complexity is available in \cite{grunwald}.

\section{Information content and intrinsic meaning}

As an attempt to define lack of meaning, think of a single bit. A single bit does not carry any information, and so it cannot but be meaningless if there is no recipient to interpret it as something richer. The Shannon entropy of a single bit is 0 because one cannot establish a communication channel--no message can be encoded or transmitted with a single letter of the alphabet. It is intrinsically meaningless because there is no context. The same is true for a string of $n$ identical bits (either 1s or 0s). To give it a meaning one would likely be forced to make an external interpretation, because even if it carries a message it cannot be intrinsically very rich, simply because it cannot carry much information. In both cases Shannon's entropy and the algorithmic complexity of such a string is very low. \\[-20pt]

\begin{center}
\colorbox{yellow!10}{\fbox{
\begin{minipage}[t]{.47\textwidth}
\textbf{Box 2. Algorithmic complexity} is the length in bits of the shortest program producing a string $s$ when running a program $p$ on a universal Turing machine $U$ upon halting. One refers to $C(s)$ as the algorithmic complexity of $s$. Formally, $C(s)=\min{|p| : U(p)=s}$ where $|p|$ is the length of $p$ measured in bits. Algorithmic complexity formalizes the concept of simplicity versus complexity. For an introduction to AIT, please see \cite{calude,li}.
 \end{minipage}\text{ }}}
\end{center}

\text{\\[10pt]}

At the other extreme, a random string cannot be usually considered meaningful. What one can say with certainty is that something meaningful should therefore lie between these two extremes: \emph{no information} (trivial) or \emph{complete nonsense} (random). Algorithmic complexity (see Box 2) opposes what is simple to what is complex or random and can be thought of as a first approximation to meaning, as opposing what is meaningful in some rough sense to what is meaningless. However, algorithmic complexity alone does not suffice to define meaning because it cannot properly oppose complexity to randomness, only simplicity to complexity or randomness. This is because algorithmic complexity associates randomness with the highest level of complexity. But using Bennett's logical depth \cite{bennett} (also rooted on algorithmic complexity) we are able to distinguish between something that looks organized and something that looks random or trivial by introducing time (a parameter that seems unavoidable in the unfolding of a message, which makes it reasonable to associate this measure with a measure of \emph{physical complexity} as Bennett himself suggests \cite{bennett2}). This connection to Bennett's logical depth means that a message cannot be instantaneously meaningful if it is not in a proper context.\\[-20pt]

\begin{center}
\colorbox{yellow!10}{\fbox{
 \begin{minipage}[t]{.47\textwidth}
\textbf{Box 3. Conditional algorithmic complexity} is defined \cite{li} as the shortest program $p$, for which the universal Turing machine $U$ outputs $s$ given $x$, that is $C(s|x)=\min{|p| : U(x,p)=s}$. Notice that the definition of algorithmic complexity in Box 2 is a special case of the conditional one when there is no $x$, i.e. $C(s)=C(s|\epsilon)$, that is the algorithmic complexity of $s$ given no other information.
 \end{minipage}\text{ }}}
\end{center}

It is generally accepted that meaning is imparted by the observer, that it is the interpretation of the message by its recipient. In order for the information conveyed to have any semantic value, it must in some manner add to the knowledge of the receiver. I claim that when it comes to mapping meaning onto information content, these semantic properties can be accounted for using concepts such as conditional algorithmic complexity (see Box 3) and logical depth (see Box 4). 

For example, conditional algorithmic complexity (see Box 3) can define a distance measure between messages. It is clear from the conditional definition that even if a string $s$ is meaningless, the complexity of $s$ given $s$ is very low because the length of the shortest program producing $s$ with input $s$ is the shortest possible.

\section{Meaning is logically deep}

As is well known, the problem with meaning is that it is highly dependent on the recipient and its interpretation. Connecting meaning to the concept of logical depth has the advantage of taking into account the story and context of a message, and therefore of potentially accounting for the likely recipient's interpretation. A meaningful message (short or long) contains a long computational history when taken together with the associated computation, otherwise it has little or no meaning. Hence the pertinence of the introduction of logical depth. \\[-20pt]

\begin{center}
\colorbox{yellow!10}{\fbox{
 \begin{minipage}[t]{.47\textwidth}
 \textbf{Box 4. Bennett's logical depth} is defined \cite{bennett} as the execution time required to generate a string by a near-incompressible program, i.e. one not produced by a significantly shorter program. Logically deep objects contain internal evidence of having been the result of a long computation and satisfy a slow-growth law (by definition).
 \end{minipage}\text{ }}}
\end{center}
\text{\\[10pt]}

Think of a winning number in a lottery. The number by itself may be meaningless for a recipient, but if two parties had shared information on how to interpret it, the information shared beforehand becomes part of the computational history and as such not unrelated to the subsequent message. The only way to interpret a number as being the winning number of a lottery is to have a story, not just a story that relates the number to a process, but one that narrates the process itself. Since winning a prize is no longer a matter of apparent chance but has to do with the release of information (both the number and the interpretation of the number), it is therefore not the number alone that represents the content and meaning of the message (the number), but the story attached to it. 

There are also messages that contain the story in themselves. If instead of a given number one substitutes the interpretation of such a number, the message can be considered meaningful in isolation. But both cases have the same logical depth, as they have the same output and computing time and are the result of the same history (even if in the first case such a history may be rendered in two separate steps) and origin, hence the definition seems robust enough.

 If a Turing machine randomly performs a lot of work when provided with a random input, making it look \emph{meaningful}, it may seem that our approach is not robust enough, since something that's taken to be meaningful is actually just a random computation. There are, however, two possible answers to this objection: On the one hand, the probability of a machine undertaking a long computation by chance is very low. Calude and Stay \cite{calude} prove that of machines that halt, most will halt after a few steps. This happens for most strings, meaning that most messages are meaningless if both the message and the computation do not somehow \emph{resonate} with each other, which recalls an intuitive requirement for considering something a meaningful message. On the other hand, also based on the results of Calude and Stay, algorithmic probability almost guarantees the non-existence of \emph{Rube Goldberg} machines (amusing machines that do a lot of work in order to accomplish a trivial task), which implies that the amount of work performed  (the number of steps and output length) by a Turing machine that halts will be proportional to the information content of the input message.

\section{An algorithmically robust definition of meaning}

Algorithmic probability, as defined by Solomonoff \cite{solomonoff} and Levin \cite{levin} (see Box 5) indicates that every outcome is likely to be produced by the shortest program(s) producing that outcome. In other words, meaningful messages would have little chance of being interpreted as such by chance.\\[-15pt]

\begin{center}
\colorbox{yellow!10}{\fbox{
 \begin{minipage}[t]{.47\textwidth}
 \textbf{Box 5. Algorithmic probability:} The algorithmic probability of a string $s$ is the probability of producing $s$ with a random program $p$ (a sequence of fair coin flip inputs) when running on a universal prefix-free Turing machine \cite{levin}. That is, a machine for which a valid program is never the beginning of any other program, so that one can define a convergent probability the sum of which is at most 1. Formally, $m(s) = \Sigma_{p : U(p) = s} 2^{-|p|}$, i.e. the sum over all the programs for which $U$ with $p$ outputs the string $s$ and halts.
\end{minipage}\text{ }}}
\end{center}\text{\\[10pt]}

It is algorithmic probability that would account for the robustness of this algorithmic approach to meaning. The meaning of a message only makes sense for particular recipients (not for any random ones). A message that has meaning for someone may not have meaning for someone else-- just the kind of property one would desire in a concept entailing the meaning of \emph{meaning}. This is what happens when some machines react to a \emph{meaningful} input rather than to a \emph{random} one. Algorithmic probability guarantees that most machines will halt almost immediately with no computational history for a given message. In other words, there is a correspondence between a meaningful input, computation time and a structured outcome, given the connection between algorithmic probability and algorithmic complexity (see Box 6).\\[-15pt]

\begin{center}
\colorbox{yellow!10}{\fbox{
 \begin{minipage}[t]{.47\textwidth}
 \textbf{Box 6. The coding theorem} is a theorem connecting algorithmic probability to algorithmic complexity. Algorithmic probability is related to algorithmic complexity in that $m(s)$ is at least the maximum term in the summation of programs given that it is the shortest program that has the greater weight in the summation of the fractions defining $m(s)$. Formally, the theorem states that the following relation holds: $-\log_2\normalsize{ }m(s) = C(s) + O(1)$. For technical details see \cite{calude}.
\end{minipage}\text{ }}}
\end{center}\text{\\[10pt]}

\section{Finite randomness (just like meaning) is in the eye of the beholder}
 
In a move that parallels the mistaken use and overuse of Shannon's measure as a measure of complexity, the notion of complexity is frequently associated, in the field of complex systems, with the number of interacting elements or the number of layers of a system. Researchers who make such an association should continue using Shannon's entropy since it quantifies the distribution of elements, but they should also be aware that they are not measuring the complexity of a system or object, but rather its diversity, which may be a different thing altogether (despite being roughly related). 

As has been shown by Stephen Wolfram, it is not always the case that the greater the number of elements the greater the complexity, nor is it the case that a greater number of layers or interactions make for greater complexity, for the simplest computing systems are capable of the greatest apparent complexity \cite{nks}.

 The theory of algorithmic randomness does not guarantee that a string of finite length cannot be algorithmically compressed. Nonetheless, any string is guaranteed to occur as a substring (with equal probability) in any algorithmically random infinite sequence. But this has to do with the semantic value of AIT, given that a finite string has meaning only in a particular context, as a substring of a larger, potentially longer and essentially different string. Therefore, one can declare a string to be random-looking only as long as it does not appear as a substring embedded in another finite or infinite string. 

A string may be \emph{random} at the scale of an infinite sequence, but since all possible strings are contained in an infinite random sequence, random blocks of strings may not always look random. For example, consider the following pseudo-randomly generated string (generated with \emph{Mathematica's} default random number generator: 1, 0, 0, 0, 0, 0, 1, 1, 0, 0, 0, 1, 1, 1, 0, 0, 1, 0, 1, 1, 1, 1, 1, 1, 1, 1, 1, 1, 1, 0, 0, 0, 0, 0, 0, 0, 0, 0, 0, 0, 0, 0, 0, 0, 0, 0, 1, 0, 1, 1, 1, 0, 0, 0, 1, 0, 0, 0, 0, 0, 1, 0, 0, 0, 1, 1, 0, 0, 0, 1, 1, 1, 0, 0, 1, 1, 1, 1, 1, 1, 1, 1, 1, 0, 0, 1, 1, 1, 0, 0, 1, 0, 1, 1, 0, 1, 0, 0, 0, 1. This string looks quite random when taken as a whole, but it actually contains several blocks that do not look random at all (e.g. 1, 1, 1, 1, 1, 1, 1, 1, 1, 1, 1 from position 19 to position 29; or 0, 0, 0, 0, 0, 0, 0, 0, 0, 0, 0, 0, 0, 0, 0, 0, 0 from position 30 to position 46) if isolated. 

On the other hand random-looking strings can always be part of larger non-random sequences. A string with period 10 (and therefore compressible to about a tenth of its size): 0, 1, 1, 1, 0, 0, 1, 0, 1, 1, 0, 0, 1, 1, 1, 0, 0, 1, 0, 1, 1, 0, 0, 1, 1, 1, 0, 0, 1, 0, 1, 1, 0, 0, 1, 1, 1, 0, 0, 1, 0, 1, 1, 0, 0, 1, 1, 1, 0, 0, 1, 0, 1, 1, 0, 0, 1, 1, 1, 0, 0, 1, 0, 1, 1, 0, 0, 1, 1, 1, 0, 0, 1, 0, 1, 1, 0, 0, 1, 1, 1, 0, 0, 1, 0, 1, 1, 0, 0, 1, 1, 1, 0, 0, 1, 0, 1, 1, 0, 0, 1, 1, 1, 0, 0, 1, 0, 1, 1, 0; contains, if isolated, a 'random-looking' string 0, 1, 1, 1, 0, 0, 1, 0, 1, 1, 0.

 The same goes for a message. A message can be contained in a longer message, and actually its interpretation may change from one context to another. From the perspective of algorithmic complexity (when interpreted in terms of meaning) this property can be seen as an indication that there cannot be an ultimate interpretation of the meaning of a message-- yet another desirable feature of algorithmic complexity that mirrors our intuitive sense of the concept of \emph{meaning}.\\[-15pt]

\begin{center}
\colorbox{yellow!10}{\fbox{
 \begin{minipage}[t]{.47\textwidth}
 \textbf{Box 7. The Halting problem and Chaitin's $\Omega$:} As widely known, the Halting problem for Turing machines is the problem of deciding whether an arbitrary Turing machine $T$ eventually halts on an arbitrary input $s$. Halting computations can be recognized by simply running them for the time they take to halt. The problem is to detect non-halting programs, about which one cannot know if the program will run forever or will eventually halt. An elegant and concise representation of the halting problem is Chaitin's irrational number $\Omega$ \cite{chaitin}, defined as the halting probability of a universal computer programmed by coin tossing. Formally, $0 < \Omega = \sum_{\normalsize{p\textbf{ }halts}} 2^{-|p|} < 1$ with $|p|$ the size of $p$ in bits. Chaitin's $\Omega$ number is the halting probability of a universal (prefix-free) Turing machine running a random program (a sequence of fair coin flip bits taken as a program).
 \end{minipage}\text{ }}}
\end{center}\text{\\[10pt]}

At the other extreme, in this algorithmic context, there is Chaitin's $\Omega$ number \cite{chaitin} that may be regarded as entailing the greatest possible meaning because it encodes all possible messages in the form of answers to all possible questions encoded by Turing machines. That Chaitin's $\Omega$ is in practice inaccessible seems desirable if we are to avoid a contradiction in the concept of meaning, and also in light of the fact that one cannot expect to encode all meanings in a single message. In other words, complete meaningfulness or the apprehension of all possible meanings is unattainable via this algorithmic approach, as it is ultimately uncomputable. Which is just what one would intuitively expect to be the case with \emph{meaning} in the broadest sense.

 \section{Towards a philosophical agenda}

The previous discussion sketches a possible agenda for a philosophy of information in the context of the current state of the theory of algorithmic information. Focusing more on the core of the theory itself, there are several directions that a deeper exploration of the foundations of algorithmic information might take. 

There is, for example, Levin's contribution \cite{levin} to algorithmic complexity in the form of the eponymous semi-measure, motivated by a desire to fundamentally amend Kolmogorov's plain definition of complexity in light of the realization that information should follow a law of non-growth conservation. This is an apparently different motivation from the one behind Gregory Chaitin's definition of algorithmic complexity in its prefix-free version. 

There are also laws of symmetry and mutual information discovered by G\'acs \cite{gacs}, and Li and Vit\'anyi \cite{li}, for example, which remain to be explored, fully understood and philosophically dissected. 

Furthermore there are the subtle but important differences involved in capturing organization in information through the use of algorithmic randomness versus doing so using Bennett's logical depth \cite{bennett}, a matter brought to our attention in, for example \cite{delahaye}. The motivation behind Bennett's formulation of his concept of logical depth was to capture the notion of complexity by taking into account the history of an object. It had, in our algorithmic approach to information, the important consequence of classifying intuitively shallow physical objects as objects deprived of meaning. 

There is also the question of the dependence of the definitions on the context in which meaning is evaluated (the choice of universal Turing machine), up to an additive constant, a question which has recently been addressed in \cite{delahayezenil,delahayezenil2}, who propose that reasonable choices of computational formalisms actually lead to reasonable evaluations of complexity \cite{delahayezeniltow}. This is the problem of finding a stable framework for a robust enough evaluation of information content.

 In \cite{floridi2}, van Benthem highlights an issue of great philosophical interest when he expresses a desire to understand the \emph{unreasonable effectiveness} (a phrase he claims is borrowed) of quantitative information theories. 

Paradoxically, my concern would be with the \emph{unreasonable ineffectiveness} of qualitative information theories, notably algorithmic complexity, given that it is the \emph{unreasonable effectiveness} of quantitative information theories, notably Shannon's notion of information entropy, that has mistakenly led researchers to use it in frameworks in which a true (and universal) measure of complexity is needed. The connections between Shannon's entropy and Kolmogorov complexity are investigated in detail in \cite{grunwald}. 

It is the ineffectiveness of algorithmic complexity that imbues information content with its deepest character, given that its full characterization cannot effectively be achieved even if it can be precisely defined. Hence Van Benthem opens up a rich vein, this discussion being potentially fruitful and of great interest even if it has hitherto been largely ignored. I think the provisional formulations of the laws of information, together with their underlying motivations should be a central part of the discussion, if not the main focus of a semantic approach to information which is more integrally rooted in current mathematical developments. 

It may be objected that a semantic approach to the study of information should not be reduced to the algorithmic, which is sometimes considered a syntactic-digital view of information. That, however, would be rather an odd objection, given that most texts on information start with Shannon's information theory, without however taking the next natural step and undertaking a discussion of the current state of information as exemplified by AIT. So either the philosophy of information ought to take a completely different path from Shannon's, which inevitably leads to the current state of algorithmic information and prompts deeper exploration of it, or else it should steer clear of the algorithmic side as being separate from and alien to it\footnote{Pieter Adriaans has presented similar arguments \cite{adriaans} in relation to the often mistaken tendency of semantic approaches to information to largely ignore the theory of algorithmic information.}. In other words, I don't find it consistent to cover Shannon's work while leaving out all further developments of the field by, among others, Kolmogorov, Chaitin, Solomonoff, Levin, Bennett, G\'acs and Landauer. As I have pointed out in the previous section, there is a legitimate agenda concerning what some may call the syntactic-mechanistic branch of the study of Information, which, paradoxically, I think is the most interesting and fruitful part of the semantic investigation, and one that mainstream Philosophy of Information has traditionally steered clear of for the most part, a few exceptions notwithstanding\cite{hermes,mcallister}. 

No account of what information might be can be considered complete without taking into account the interpretations of quantum information. One relevant issue has been raised by Wheeler \cite{wheeler}, though perhaps with reference to a different scale, and that is whether an observer is necessary for information to exist and for an observation to have meaning. There does not exist a universally accepted interpretation of quantum mechanics, although the so-called \emph{Copenhagen Interpretation} is considered the mainstream one. Discussions about the meaning of quantum mechanics and its implications do not, however, lead to a consensus. It is beyond the scope of this paper to further discuss the quantum approach other than to point out its pertinence in an encompassing discussion. See for instance \cite{cabello}.

\subsection{Basics to agree upon}

We should agree upon fundamental properties of information derived from the current state of AIT that, as I've argued, can serve as a basis for a mathematical framework in a philosophical discussion. Although this is not the place to discuss the several results of the theory of algorithmic complexity, here is a non-exhaustive list of some of the points to be agreed upon, together with the claims that meaning can be captured and at least some of its properties studied. The list, in a certain order of logical derivation, is:

\begin{myitemize}
\item The basic unit of information is the bit, but information remains subjective (Shannon \cite{shannon}).
\item Shannon's information measure cannot capture content, organization or meaning as it is neither a measure of information content nor of complexity.
\item Algorithmic complexity is an objective and universal framework for capturing structure and randomness.
\item Randomness implies the impossibility of information extraction (Chaitin \cite{chaitin}).
\item Shallowness is meaningless (Kolmogorov \cite{kolmogorov}).
\item Randomness is meaningless (Bennett \cite{bennett}).
\item There are strong connections between logical and thermodynamic (ir)reversibility to be explored (Bennett \cite{bennett3}, Fredkin \cite{fredkin}, Toffoli \cite{fredkin}).
\item Information can be transformed into energy and energy into information (Landauer \cite{landauer}, Bennett \cite{bennett}).
\item Information follows fundamental laws: symmetry, non-growth, mutual information and (ir)reversibility (G\'acs \cite{gacs}, Zvonkin \cite{zvonkin}, Levin \cite{levin}, Bennett \cite{bennett3}, Landauer \cite{landauer}).
\item Matter and information are deeply connected, although it is an open question which of the two is more fundamental(Wheeler \cite{wheeler}, Feynman \cite{feynman}, Bennett \cite{bennett3}, Landauer \cite{landauer}, Fredkin \cite{fredkin}, Wolfram \cite{nks}).
\end{myitemize}

Information has also begun playing a major role in the interpretation of quantum mechanics \cite{lloyd,cabello} and is assuming foundational status in some models of modern physics \cite{svozil} as it did in classical physics, notably in thermodynamics, and more recently in cosmology \cite{bekenstein}. We shall further survey the listed properties and the connections of information to physics in a follow-up to this paper.

\section{Concluding remarks}
 
A common language and a commonly agreed upon formal framework seem to be necessary. I've claimed that algorithmic information is suitable for defining individual information content and for providing a characterization of the concept of meaning in terms of logical depth and algorithmic probability. This rather formal computational characterization does not mean that a discussion of algorithmic information would be deprived of legitimate philosophical interest. 

I have briefly drawn attention to and discussed some of the questions germane to a philosophy of algorithmic information in connection to a semantic definition of meaning. I've argued that algorithmic complexity and algorithmic probability taken together with Bennett's logical depth can constitute an appropriate computational framework within which to discuss information content and meaning. Nor does the fact that meaning can be fully formalized mean either that it will lose its most valuable characteristics, such as subjectivity with respect to a recipient. Such a subjective and rich dimension can be computationally grasped as proposed herein, and can constitute another point of departure for an organized philosophical discussion accounting for and covering a field that can no longer be ignored in philosophical discussions of information.

\subsection*{Acknowledgments}

The author would like to thank Peter Boltuc for his suggestions, and Selmer Bringsjord for his comments, objections and valuable views on information from his more purely logical perspective. Selmer and I may not agree but we agree to disagree.

\end{multicols}
 
 \end{document}